\documentclass{moriond}

\def\Journal#1#2#3#4{{#1} {\bf #2}, #3 (#4)}

\def\JCAP{\em JCAP}

\newcommand{\mymatrix}[1]{\mathbf{#1}}
\newcommand{\myset}[1]{{#1}}
\newcommand{\polarization}{}
\newcommand{\myvector}[1]{\mathbf{#1}}
\makeatletter
\newcommand\notsotiny{\@setfontsize\notsotiny\@vipt\@viipt}
\makeatother

\usepackage{amsmath}

\usepackage{multirow}

\def\be{\begin{equation}}
\def\ee{\end{equation}}
\def\bea{\begin{eqnarray}}
\def\eea{\end{eqnarray}}

\newcommand{\Photo}{}

\begin{document}
\vspace*{4cm}
\title{Diffuse polarized foregrounds from component separation with QUIJOTE-MFI}

\author{E. de la Hoz on behalf of the QUIJOTE collaboration}

\address{Instituto de F\'isica de Cantabria (CSIC-Universidad de Cantabria),\\ 
Avda. de los Castros s/n, E-39005 Santander, Spain}

\maketitle\abstracts{
Polarized component maps in the Northern Sky are  derived from the  QUIJOTE-MFI wide survey data at 11 and 13 GHz, the WMAP K and Ka bands and all Planck polarized channels using the parametric component separation method \texttt{B-SeCRET}. The addition of QUIJOTE-MFI data significantly improves the uncertainty in the parameter estimation of the low frequency dominant foreground, in particular the estimation of the synchrotron spectral index. We find statistically significant spatial variability across the sky. A power law model of the synchrotron emission provides a good fit of the data outside the galactic plane but fails to track the complexity of this region. Moreover, when we assume a synchrotron model  with uniform curvature we find, in the 95\%  confidence region, a non-zero $c_s$ value. However, there is not sufficient statistical significance to determine which model is favoured.}

\section{Introduction}

One of the major milestones of the CMB scientific community is the  detection of primordial $B$-modes. Such a detection would constitute a compelling evidence of the existence of an inflationary phase at the beginning of the Universe. However, this signal is hidden among other $B$-mode sources e.g., astrophysical foregrounds, lensed $E$-modes, instrumental noise, etc. A comprehensive knowledge of astrophysical foregrounds is crucial to recover this faint signal from the measured sky data. Here, we have characterized diffuse polarized foregrounds performing component separation analyses with the new wide survey maps from the Q-U-I JOint TEnerife  Multi-Frequency Instrument (QUIJOTE-MFI) experiment  along with data from WMAP and Planck. The inclusion of QUIJOTE-MFI data in the analysis helps significantly with the characterization of the synchrotron emission \cite{CS_pol}.

\section{Component Separation Methodology}

The component separation analyses are performed using the Bayesian-Separation of Components and Residuals Estimate Tool (\texttt{B-SeCRET}) \cite{BSeCRET}. \texttt{B-SeCRET} is a parametric Bayesian pixel-based maximum-likelihood 
pipeline which uses an MCMC to sample from the posterior distribution:
\begin{equation}
    \mathcal{P}(\myset{\theta}\polarization_{p}|\myvector{d}\polarization_{p}) \propto \mathcal{P}(\myvector{d}\polarization_{p}|\myset{\theta}\polarization_{p}) \mathcal{P}(\myset{\theta}\polarization_{p}) \, .
    \label{eq:posterior}
\end{equation}
where $\myset{\theta}\polarization_{p}$ and $\myvector{d}\polarization_{p}$ are the model parameters and the signal at the pixel $p$, $\mathcal{P}(\myset{\theta}\polarization_{p})$ is the prior information and,  $\mathcal{P}(\myvector{d}\polarization_{p}|\myset{\theta}\polarization_{p})$ is  the likelihood:
\begin{equation}
    \mathcal{P}(\myvector{d}\polarization_{p}|\myset{\theta}\polarization_{p}) = \frac{\exp\left(-\frac{1}{2}\left(\myvector{d}\polarization_{p}-\myvector{S}_{p}\polarization\right)^{T}\mymatrix{C}^{-1}\left(\myvector{d}\polarization_{p}-\myvector{S}_{p}\polarization\right)\right)}{\sqrt{(2\pi)^{N}\det (\mymatrix{C})}} \, ,
    \label{eq:likelihood}
\end{equation}
where $\mymatrix{C}$ is the noise covariance matrix, $N$ is the number of elements in the $\myvector{d}_{p}$ array, and $\myvector{S}_{p}$ is the parametric model considered. The models and priors used are shown in Table~\ref{tab:models}.

\begin{table}[t]
\caption[]{Parametric models (antenna units) and parameters priors used to fit the sky polarized components.}
\label{tab:models}
\vspace{0.4cm}
\begin{center}
\begin{tabular}{|ccc|}
\hline
Component & Model &  Prior \\
\hline
CMB & $\begin{bmatrix}
c^{\textrm{\notsotiny Q}}\\
c^{\textrm{\notsotiny U}}\end{bmatrix}\dfrac{x^2e^x}{(e^x-1)^2}$ & - \\[.6cm]
\multirow{6}{*}{Synchrotron}  & 
\multirow{5}{*}{
$\begin{cases}
\begin{bmatrix}
a_s^{\textrm{\notsotiny Q}}  \\
a_s^{\textrm{\notsotiny U}}
\end{bmatrix}\left(\dfrac{\nu}{\nu_s}\right)^{\beta_{s}} \\
\\
\begin{bmatrix}
a_s^{\textrm{\notsotiny Q}}  \\
a_s^{\textrm{\notsotiny U}}
\end{bmatrix}\left(\dfrac{\nu}{\nu_s}\right)^{\beta_{s}+c_s\left(\nu/\nu_{s}\right)}
\end{cases}$
 }
& \multirow{5}{*}{
$
\left.
\begin{matrix}
\\
{\beta}_{s} \in \mathcal{N}(-3.1, 0.3)\\
\\[.3cm]
\begin{cases}
{\beta_{s}} \in \mathcal{N}(-3.1, 0.3)\\
c_{s} \in \mathcal{N}(0, 0.1)
\end{cases}
\end{matrix}
\right.
$}\\
&&\\
&&\\
&&\\
&&\\[1cm]
Thermal dust & $\begin{bmatrix}
a_d^{\textrm{\notsotiny Q}}  \\
a_d^{\textrm{\notsotiny U}}
\end{bmatrix}\left(\dfrac{\nu}{\nu_d}\right)^{\beta_{d}-2}
\dfrac{\rm{B}\left(\nu,T_{d}\right)}{\rm{B}\left(\nu_d,T_{d}\right)}$ & $\begin{cases}
{\beta_{d}} \in \mathcal{N}(1.55, 0.1)\\
{T_{d}} \in \mathcal{N}(21, 3)
\end{cases}$ \\[.2cm]
\hline
\end{tabular}
\end{center}
\end{table}

\section{Data}

The QUIJOTE-MFI \cite{QUIJOTE-MFI} instrument is a polarimetric CMB experiment in the 10-20GHz microwave range located at the Teide observatory. The instrument consists of 4 horns, each of which has 8 independent channels. Combining all the data, the QUIJOTE-MFI provides four frequency bands centred around 11, 13, 17 and 19 GHz with a bandwidth of approximately 2 GHz. The low-frequency bands (11 and 13 GHz) have an approximate angular resolution of 52 arc min while the 17 and 19 GHz channels have one of approximately 38 arc min.

In this study we have used the results from the MFI wide survey, a shallow survey where the instrument covered all the visible sky from Tenerife at elevations above 30$^{\circ}$. After removing the Radio Frequency Interference (RFI) from the geostationary satellite band and some areas with large atmospheric air mass we are left with $\sim$ 51\% of the sky. This survey provides an average sensitivity in polarization of $\sim 35$-$40 \mu$K deg$^{-1}$. In the analysis we used only the 11 and 13 GHz, since they have better sensitivity and the largest signal-to-noise ratio.

We have used the low frequency WMAP's K  (22.8 GHz) and Ka (33.1 GHz) bands \cite{WMAP} and Planck's polarization maps from Planck Release 4 (PR4) \cite{NPIPE} i.e., the low frequency instrument (LFI) 30, 44 and 70 GHz frequency maps and the high frequency instrument (HFI) 100, 143, 217 and 353 GHz maps. 
    
All the maps are smoothed to the common angular resolution
of $2^{\circ}$, and downgraded to $N_{side} = 64$. We have calculated the frequency covariance  $\mymatrix{C}$ using noise simulations specific to each instrument. In the case of QUIJOTE \cite{QUIJOTE-MFI} and Planck \cite{NPIPE} we have used the noise simulations provided by the collaboration. For WMAP, we generated a set of white noise simulations  using  the  rms  noise  per  pixel  provided  by  the WMAP collaboration \cite{WMAP}.  To account for the unequal detectors response across their bandwidth  we apply colour corrections calculated using the python code \texttt{fastcc} \cite{pipeline}.

\section{Results}

\begin{figure}
\centering
\includegraphics[width=.96\linewidth, trim={2cm 1.6cm 2.5cm 1.7cm}, clip]{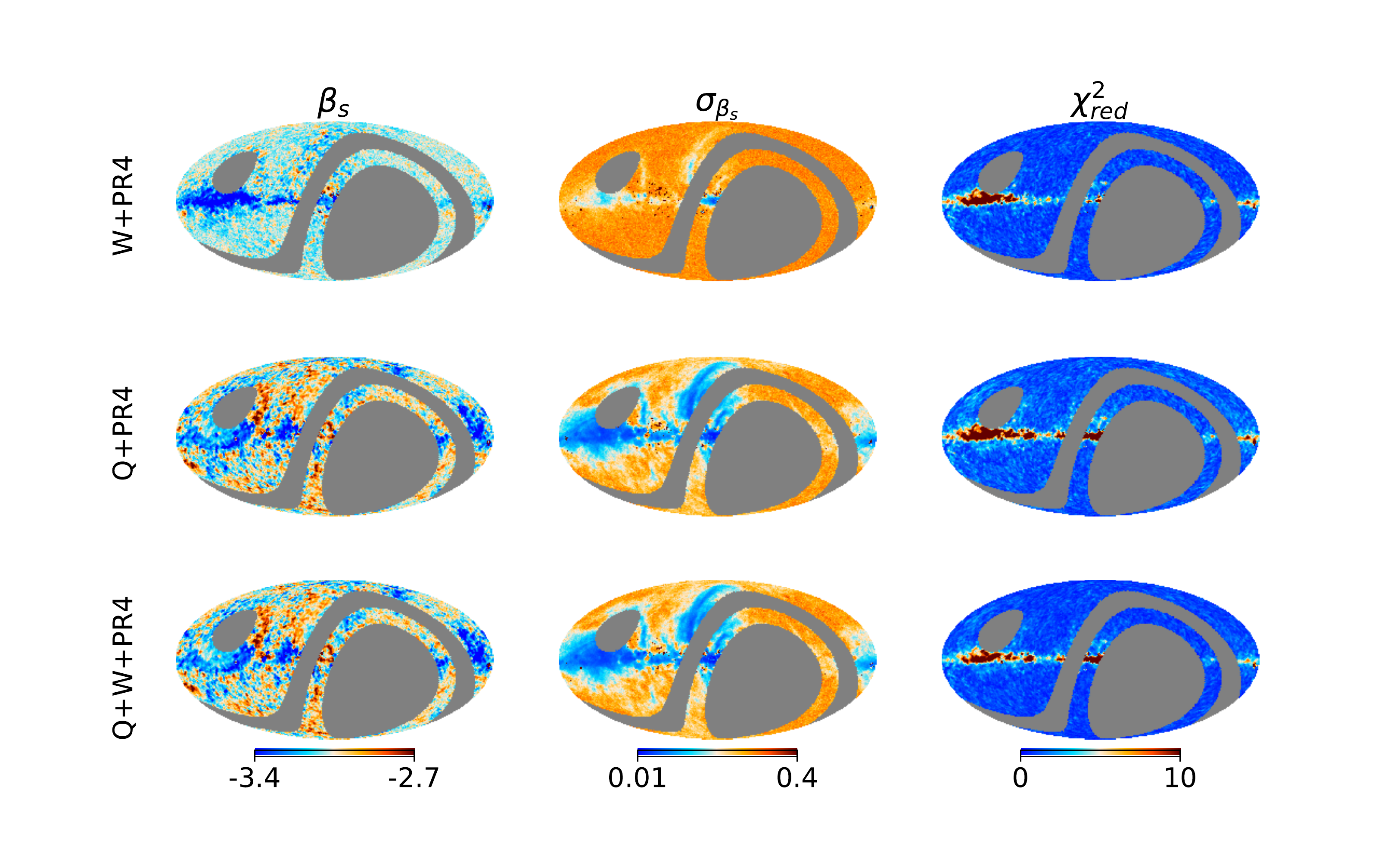}
\caption[]{$\beta_s$ (left column), $\beta_s$ uncertainty (center column) and reduced $\chi^2$ statistic obtained from the fit using a power law model using different data. WMAP and Planck data (top row), QUIJOTE and Planck data (center row), QUIJOTE, WMAP and Planck (bottom row).}
\label{fig:datasets}
\end{figure}
Fig.~\ref{fig:datasets} shows the estimated synchrotron spectral index ($\beta_s$) map and its uncertainty map obtained from the component separation analysis when the synchrotron is modeled with a power law using different combinations of the available data. The reduced $\chi^2$ statistic map is also shown. It is clear that the inclusion of QUIJOTE-MFI channels improves the characterization of $\beta_s$, specially in the regions where the synchrotron has the largest signal-to-noise-ratio. Besides, the power law model represents well the data except along the galactic disk where the physics might be more complex.

\begin{figure}
\centering
\includegraphics[width=.75\linewidth, trim={2cm .68cm 2cm 1.5cm}, clip]{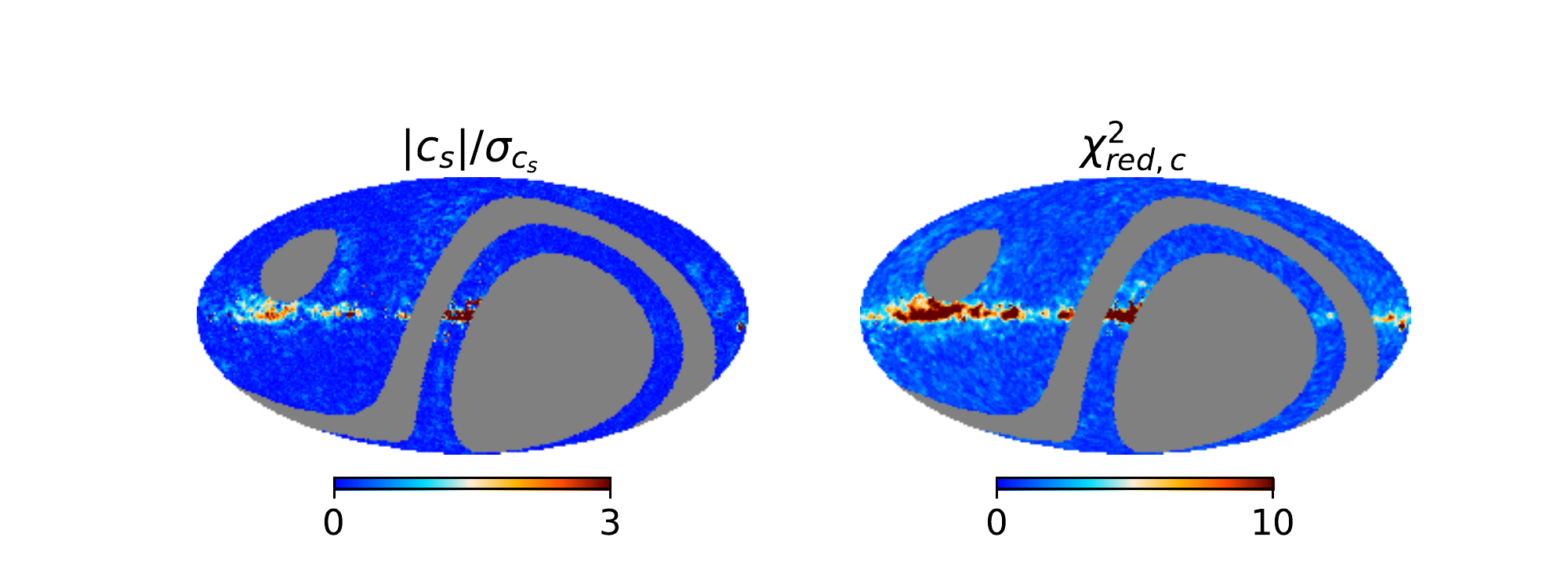}
\caption[]{Signal-to-noise map of the recovered synchrotron curvature and the reduced $\chi^2$ statistic obtained from the fit using a power law with spatially varying curvature model.}
\label{fig:pixelwise_c_s}
\end{figure}

We have also modeled the synchrotron emission as a power law with spatially varying curvature. The pixel-based analysis of the curvature shows that $c_s$ is only detected in some regions at the galactic disk where the fit is bad, see Fig.~\ref{fig:pixelwise_c_s}. 

Moreover, we considered the case where the synchrotron's curvature is uniform in a given region. We have studied four regions (Table~\ref{tab:c_s_uniform}): i) RC1 composed of the pixels whose $\chi^2_{red}$ is  within  95\%  confidence  region \footnote{Pixels whose $\chi^2$ value is lower than the critical value that satisfies that the probability of exceeding itself is less than 0.05 given a $\chi^2$-distribution with the appropriate  degrees of freedom.}, ii)  RC2  is a subset of the  RC1  pixels  that satisfy  that  the    synchrotron’s polarized intensity signal-to-noise ratio is larger than 5 and,  iii) the Haze and iv) North bubble which are two physically-defined regions studied in more detailed in a companion QUIJOTE paper \cite{Haze}. In all cases we detect a non-zero negative curvature. We found that both models, i.e., power law and power law with uniform curvature, provide a good fit given the data available. However there is not enough statistical significance to distinguish which model is better. A more thorough study is left for further work.

\begin{table}[t]
\caption[]{Estimated curvature and its uncertainty values obtained assuming that the curvature is  uniform within the region.}
\label{tab:c_s_uniform}
\vspace{0.4cm}
\begin{center}
\begin{tabular}{|cccc|}
\hline
Region  & $c^R_s$ & $\sigma_{c^R_s}$ & $\left|c^R_s\right|/\sigma_{c^R_s}$ \\
\hline
    RC1  & -0.1095 &  0.0013 & 82 \\ 
    RC2  &  -0.3107 &  0.0019 & 161 \\  
    Haze  & -0.046 & 0.010 & 4.78 \\
    North bubble  &  -0.044 & 0.007 & 6.29 \\ 
\hline
\end{tabular}
\end{center}
\end{table}

\section{Conclusions}

The inclusion of the low-frequency QUIJOTE-MFI data improves  significantly the characterization of the synchrotron model parameters, specially the synchrotron spectral index. We find statistically significant spatial variability of $\beta_s$ across the sky. We obtain that a power law and power law with curvature model well the synchrotron emission outside the Galactic plane but fails to track the complexity of this region. Also, when we assume uniform curvature we detect a non-zero negative curvature. However, our results are not robust enough to distinguish the model, either the power law or the power law with uniform curvature, which fits better the data.

\section*{Acknowledgements}

We thank the staff of the Teide Observatory for invaluable assistance in the commissioning and operation of QUIJOTE.
The QUIJOTE experiment is being developed by the Instituto de Astrofisica de Canarias (IAC), the Instituto de Fisica de Cantabria (IFCA), and the Universities of Cantabria, Manchester and Cambridge.
Partial financial support was provided by the Spanish Ministry of Science and Innovation under the projects AYA2007-68058-C03-01, AYA2007-68058-C03-02, AYA2010-21766-C03-01, AYA2010-21766-C03-02, AYA2014-60438-P, ESP2015-70646-C2-1-R, AYA2017-84185-P, ESP2017-83921-C2-1-R, AYA2017-90675-REDC (co-funded with EU FEDER funds),
PGC2018-101814-B-I00, PID2019-110610RB-C21, PID2020-120514GB-I00, IACA13-3E-2336, IACA15-BE-3707, EQC2018-004918-P, the Severo Ochoa Programs SEV-2015-0548 and CEX2019-000920-S, the Maria de Maeztu Program MDM-2017-0765, and by the Consolider-Ingenio project CSD2010-00064 (EPI: Exploring the Physics of Inflation). We acknowledge support from the ACIISI, Consejeria de Economia, Conocimiento y Empleo del Gobierno de Canarias and the European Regional Development Fund (ERDF) under grant with reference ProID2020010108.
This project has received funding from the European Union's Horizon 2020 research and innovation program under grant agreement number 687312 (RADIOFOREGROUNDS). Some of the results in this paper have been derived using the HEALPix (K.M. Górski et al., 2005, ApJ, 622, p759) package.

\end{document}